\documentclass[11pt,a4paper]{article}
\usepackage[hyperref]{acl2021}
\usepackage{times}
\usepackage{latexsym}

\usepackage{amsmath}

\usepackage{microtype}
\usepackage{todonotes}
\aclfinalcopy 

\title{Self-supervised Answer Retrieval on Clinical Notes}

\author{Paul Grundmann \hspace{1cm} Sebastian Arnold \hspace{1cm} Alexander Löser\\
  Beuth University of Applied Sciences \\
  Berlin, Germany \\
  \texttt{\{pgrundmann,sarnold,aloeser\}@beuth-hochschule.de} \\
  }

\date{}

\begin{document}
\maketitle
\begin{abstract}
Retrieving answer passages from long documents is a complex task requiring semantic understanding of both discourse and document context. We approach this challenge specifically in a clinical scenario, where doctors retrieve cohorts of patients based on diagnoses and other latent medical aspects. We introduce CAPR, a rule-based self-supervision objective for training Transformer language models for domain-specific passage matching. In addition, we contribute a novel retrieval dataset based on clinical notes to simulate this scenario on a large corpus of clinical notes. We apply our objective in four Transformer-based architectures: Contextual Document Vectors, Bi-, Poly- and Cross-encoders. From our extensive evaluation on MIMIC-III and three other healthcare datasets, we report that CAPR outperforms strong baselines in the retrieval of domain-specific passages and effectively generalizes across rule-based and human-labeled passages. This makes the model powerful especially in zero-shot scenarios where only limited training data is available.
\end{abstract}

\section{Introduction}

\begin{figure}[t]
    \centering
    \includegraphics[width=\linewidth]{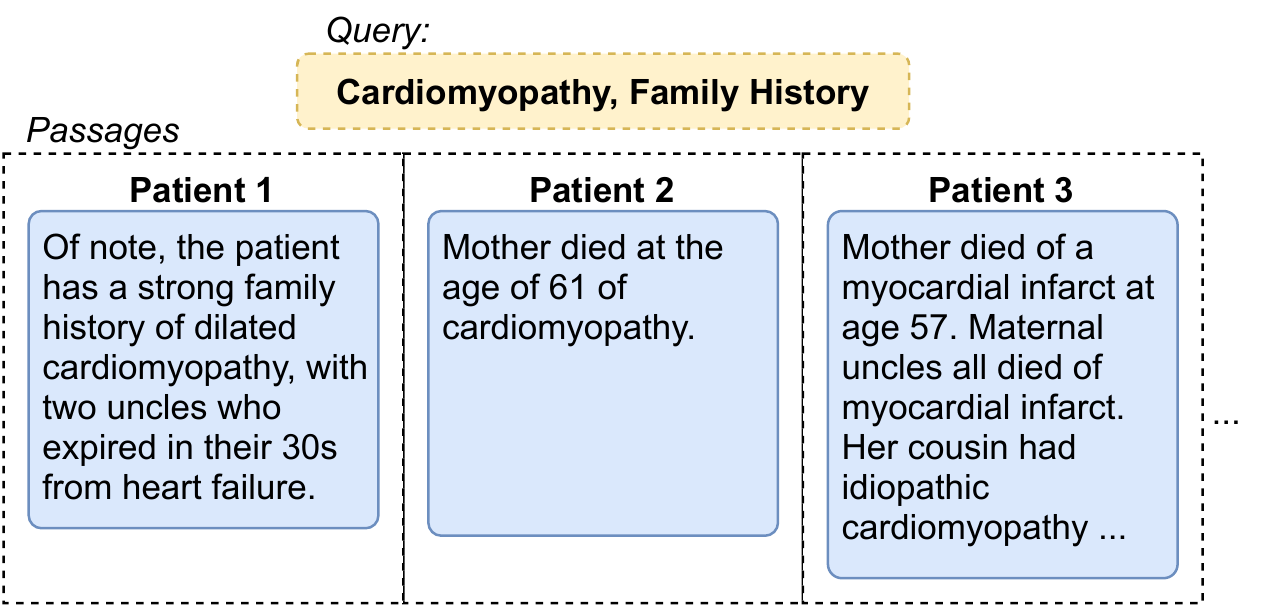}
    \caption{Application of CAPR showing three relevant retrieval results of MIMIC-III passages (blue) for a query (yellow) involving the entity ``Cardiomyopathy'' and the latent clinical aspect ``Family History''.}
    \label{fig:doc_visualization}
\end{figure}

Open-ended search intentions, such as questions about health, often lead to complex answers that span multiple sentences (see \autoref{fig:doc_visualization}). These passages are often embedded in the context of long documents, where multiple latent query-related aspects might be discussed. Retrieving such non-factoid answers from a large corpus of passages is a complex task that requires semantic understanding of discourse and document context \cite{arora2016latent, nanni2017benchmark, cohan2018discourse}. Traditional retrieval methods often assume that answer passages are independent and retrieve them using term matching \cite{o1980answer, salton1993approaches} or learn to rank them using human-annotated query--answer pairs as training data \cite{tellex2003quantitative, yang2016beyond, zhu2019hierarchical}. More recent approaches aim to find passages in the context of longer documents \cite{keikha2014retrieving, arnold2019sector} and utilize hierarchical attention \cite{zhu2019hierarchical} or latent neural representations \cite{arnold2020learning} for retrieving them with high recall.

A major challenge in this scenario is to apply neural contextual architectures for ad-hoc retrieval from large datasets with low latency. Typically, the Cross-encoder architecture focuses on the interaction between query and passage at the significant cost of computation \cite{hu2014convolutional,guo2016deep,yang2016anmm}. A more efficient solution is the Bi-encoder approach, where the query is encoded independently from the passage \cite{huang2006evaluation,shen2014learning}. This model often trades speedup for retrieval accuracy. To overcome this problem, Poly-encoders combine the efficiency of Bi-encoders with the effectiveness from additional interaction features \cite{humeau2020polyencoders}.

\paragraph{Our contribution.} 
We approach the task of retrieving answer passages for doctors, who search for clinical manifestations in thousands of health records. Our setup resembles the clinical back\-ground--fore\-ground scenario \cite{cheng2004study}, where a query focuses on an \emph{entity} (e.g., a disease or health problem) and a specific \emph{aspect} of the entity (e.g., its anamnesis or pathology, see \autoref{fig:doc_visualization}). This is especially challenging because in this domain, there is no task-specific training data available. The main contributions of this paper include:
\begin{itemize} 
\item We introduce Clinical Answer Passage Retrieval\footnote{Code and evaluation data available at [ANONYMIZED]} (CAPR), a novel training objective that allows self-supervised fine-tuning of language models, such as BioBERT \cite{lee2019biobert}. We apply CAPR in four recent Transformer-based architectures: Contextual Discourse Vectors (CDV), Cross-encoder, Bi-encoder and Poly-encoder.
\item Our method requires no external labels as training data. Instead, we utilize features extracted from Named Entity Recognition (NER) and corpus-specific rules. This enables us to train self-supervised, using in-domain unlabeled clinical data.
\item We apply CAPR in a clinical answer passage retrieval scenario on 7.4k clinical notes from MIMIC-III and evaluate it on three additional human annotated healthcare datasets: WikiSectionQA, MedQuAD and HealthQA.
\item Our experiments show that CAPR significantly increases cross-domain retrieval performance with up to 72.9\% Recall@1 on MIMIC-III. This is a 15.7 points increase compared to fine-tuning CDV+BioBERT on the same task.
\end{itemize}
While Poly-encoders have shown significant improvements in document retrieval, this paper is the first to utilize this architecture in a self-supervised answer retrieval task. 



This paper is structured as follows: We overview related work in Section \ref{sec:relatedwork}. We introduce our approach in Section \ref{sec:method}. In Section \ref{sec:evaluation} we present our experimental evaluation and a deep error analysis. In Section \ref{sec:conclusion}, we draw a conclusion.

\section{Related Work}
\label{sec:relatedwork}
Traditionally, passage retrieval is based on overlapping terms in queries and passages \cite{o1980answer,salton1993approaches} with term weighting strategies such as TF-IDF \cite{sparck1972statistical} or BM25 \cite{robertson1995okapi}. This exact match strategy often suffers from low recall. To overcome this problem, neural approaches match between query and passage with a similarity function learned on contextual features, such as feed-forward networks \cite{huang2013learning}, convolutional neural networks \cite{hu2014convolutional} or  word count histograms \cite{guo2016deep}. More recently, \citet{zhu2019hierarchical} (HAR) used attention and GloVe embeddings \cite{pennington2014glove} to capture local semantic features between a given passage and query. 

\paragraph{Transformer-based passage retrieval.} \citet{humeau2020polyencoders} apply three different Transformer-based architectures to the passage retrieval task, which offer a different level of coupling between query and passage representations. \emph{Bi-encoders} initially learn individual networks for query or passage and later a matching function for these independent representations. \emph{Cross-encoders} apply full self-attention between query and passage, allowing for direct interaction between words or phrases. \citet{humeau2020polyencoders} propose \emph{Poly-encoder} as a compromise between the high computation cost of Cross-encoder and the efficiency of the Bi-encoder to gain better performance. Furthermore, \citet{khattab-2020-colbert} show that a late interaction in the form of a fine-grained similarity matching between query- and document representation can improve over regular Bi-encoders.

\paragraph{Contextual discourse representation.} \citet{arnold2020learning} propose a dual-encoder architecture that embeds discourse information from the entire document into passage representations. Their \emph{Contextual Discourse Vectors} (CDV) model utilises a bidirectional LSTM \cite{hochreiter1997long} to encode a sequence of sentence representations generated by either GloVe, FastText \cite{bojanowski2017enriching} or BioBERT embeddings \cite{lee2019biobert}. CDV does not focus on term interaction, but instead projects query entities and aspects into pre-trained embedding spaces.

We build upon the CDV query model and propose to fine-tune a Transformer model within a self-supervised scenario. We aim to compare our approach between Bi-, Poly- and Cross-encoder architectures in order to show the applicability of these models to a zero-shot transfer learning task in the clinical domain.

\section{Transformer-based Answer Retrieval}
\label{sec:method}

We introduce Clinical Answer Passage Retrieval (CAPR), a training objective for self-supervised passage retrieval. CAPR enables the fine-tuning of passage retrieval models on in-domain clinical data without requiring external labels or answers. To show the applicability of this objective, we adapt four transformer-based model architectures for this task and compare their properties.

\subsection{Training Objective}
We aim to tune the model for two semantic question dimensions that are based on the clinical background-foreground principle \cite{cheng2004study} and have been adapted for application in clinical decision support by \citet{arnold2020learning}: \emph{Entity} describes the clinical entities that are discussed in a passage, such as diseases, procedures or medications. \emph{Aspect} describes the main clinical topic covered by the passage, e.g. ``chief complaint'', ``medical history'' or ``allergies''. Following this task definition, we represent a query as a tuple of entity and aspect and optimize our model to retrieve passages that discuss both these dimensions.

\paragraph{Label representation.}
Differently to the approach of \citet{arnold2020learning}, we use the raw string representation of entities and aspects as sparse input to our model instead of pretrained embeddings. This is necessary because in a clinical setting, we cannot rely on curated training data for training representations of ten thousands of entities and aspects. Instead, we focus on sparse labels that we can directly extract from the training documents. In addition, this approach allows the annotation of multiple entity labels per passage and sentence.

\paragraph{Pattern-based training label generation.}
In some data sources, such as Wikipedia pages, training labels are already sparsely annotated by the Wiki authors using page titles, section headings and page links \cite{arnold2019sector}. In domain-specific corpora, such as MIMIC-III, there are no passage-specific labels available. Therefore we utilize two pattern-based approaches to generate labels for self-supervision:

We utilize a medical Named Entity Recognition (NER) model to extract clinical entity mentions from the text and assign them as sparse entity training labels to each passage. We use TASTY \cite{arnold2016robust} because it provides a robust medical model with high recall based on Wikipedia. We propose to extend this step in future work using multiple NER models such as FLAIR \cite{akbik-etal-2018-contextual} or SciSpaCy \cite{neumann2019scispacy} to increase the number of recognized entities.

We detect the aspect labels by extracting them from the document structure. Doctors typically write clinical aspects as headings into the document. We extract these terms by matching the template style of the documents using regular expressions and assigning them as aspect label to the corresponding passage. We skip passages without entity or aspect label during training.

\paragraph{Sampling answers for learning to rank.}
In an answer passage retrieval setting, a single passage can be linked to multiple queries. To train the model, we randomly sample one entity and one aspect label from the training data as query and use the surrounding passage as the target passage.

Following the approach of \citet{humeau2020polyencoders}, we apply listwise learning to rank, similar to ListNet \cite{cao2007listnet}, to learn a mapping from a query representation to a matching passage. 
For each training step, we sample a mini-batch of 32 query-passage pairs. We use the attached passage as positive example and every other passage in the mini-batch as negative example. Next, we minimize the cross entropy loss $\mathcal{L}$ for all contained queries $Q = \{q_1, q_2, \dots, q_n\}$ and all passages $P = \{p_1, p_2, \dots, p_n\}$:
\begin{equation}
    \mathcal{L} = - \frac{1}{|P|} \sum_{p \in P} \sum_{q \in Q} \hat{\mathrm{y}}(q, p)\ \mathrm{log}(\mathrm{y}(q, p)) 
\end{equation}
where $\mathrm{y}(q, p)$ is the predicted relevance score for a pair of query and passage $(q,p)$ and $\hat{\mathrm{y}}(q, p)$ is the ground truth data.



\begin{figure}[t!]
    \includegraphics[width=\linewidth]{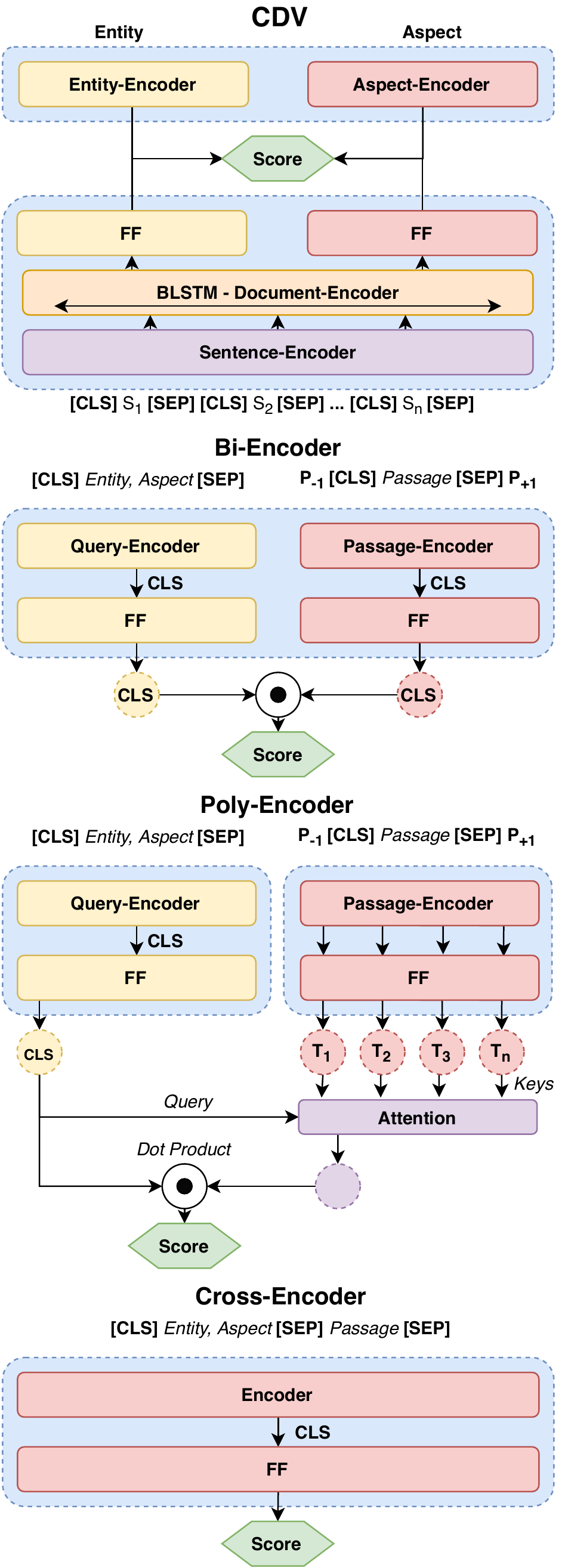}
    \caption{CDV, Bi-, Poly- and Cross-encoder models.}
    \label{fig:architecture_diagram}
\end{figure}

\subsection{Model Architectures}

\autoref{fig:architecture_diagram} shows the four model architectures we use to optimize our training objective. Next, we discuss the properties of each architecture.

\paragraph{CDV: Hierarchical Transformer+LSTM architecture.}
CDV uses fixed precomputed sentence embeddings and a bidirectional LSTM for handling long-range context. We extend this architecture to enable fine-tuning the BERT based sentence encoder jointly with the entire model. In initial experiments, this model was prone to catastrophic forgetting, therefore we propose a hierarchical fine-tuning approach.
In a first step, we train CDV with a frozen BERT layer using average pooling until convergence. Subsequently, we unfreeze BERT and use the representation of the \texttt{[CLS]} token of the last hidden layer for fine-tuning BERT jointly with the other layers.

\paragraph{Bi-encoder: Representation-based matching for fast retrieval.}
We use a Bi-encoder approach similar to \citet{humeau2020polyencoders} and train distinct query and passage encoders to learn similar representations for given query-passage pairs. In contrast to CDV, which requires pretrained encoders for the representation of the query, we input the query and the passage directly as word piece tokenized texts. This allows the model to apply full self-attention on the combination of a specific entity and aspect in the query. The Bi-encoder uses a pretrained language model such as BERT for both encoders.
Primarily, the Bi-encoder allows fast retrieval by precomputing the passage representations. We cache these representations in a FAISS vector index \cite{johnson2017billionscale} and use a k-nearest neighbor search to find the best matching passage for a given query.
Due to the high number of parameters in transformer-based language models (110M in BERT-base) and the small amount of training data, we investigate the effect of weight sharing between query and passage-encoders. This reduces the number of parameters by a factor of two. To distinguish the inputs, we mark them with \texttt{[PASSAGE]} and \texttt{[QUERY]} tokens. 
The Bi-encoder is limited by the underlying BERT model to 512 tokens of input sequence length. Because long contexts play a crucial role in clinical notes, we follow the approach of \cite{Beltagy2020Longformer} and apply the Longformer language model to encode longer sequences. 
We improve upon the BERT based version of the Bi-encoder by extending the maximum input sequence length to 4096 tokens. We concatenate surrounding passages with the input passage and use global attention. This permits the model to use parts of the document-level context for the passage representation.

Finally, we compare different pretrained language models as input encoders: BERT-base, Longformer and BioBERT.
We use CLS-pooling for both the encoded query and the context sequence to create an equally sized sequence embedding. We use the dot product between both representations as a similarity function.

\paragraph{Poly-encoder: Late interaction between query and passage.}
As a third approach, we use Poly-encoders \cite{humeau2020polyencoders} with additional attention between query and passage representations. The additional attention allows to extract specific features from the passage to enhance their representation. This adds slightly more computational complexity over the Bi-encoder, but is still less expensive than full self-attention. We define the scoring function for a given query $q$ and a passage $P = \{t_1, t_2,\dots,t_n \}$ consisting of $n$ tokens $t$ as follows:
\begin{equation}
    \begin{split}
    y &= \mathrm{softmax}(Pq^\top) \\
    s &= \mathrm{relu}(W_{\mathrm{attn}}y) \cdot q\\ 
    \end{split}
\end{equation}
where $W_{\mathrm{attn}}$ is a learned weight matrix in the attention layer and $y$ is the matrix-vector product of query- and passage-embeddings. Finally, the score $s$ is the dot product between the latent passage and query representation.
As with the Bi-Encoder, we use BioBERT as the passage and query encoder. The Poly encoder has similar problems as the Bi-encoder regarding the maximum input sequence length when combined with BERT. Therefore, we also apply the Longformer as the query and passage encoder to extend the processable input sequence length to 4096 tokens. We apply local attention to the given passage and global attention to the surrounding passages. We directly use the token embeddings as input to the attention. This adds computational complexity but improves results. For the Longformer, we apply an attention mask to the last attention layer, which masks the token embeddings that are not part of the considered passage.

\paragraph{Cross-encoder: Full query--passage interaction.}
Cross-encoder is an end-to-end trained model which applies full self-attention to the concatenated query and passage. This permits the model to focus on token-level features between query and passage. 
We concatenate query and passage and feed them directly into a single BERT encoder. However, this concatenation prohibits caching of query or passage representations, which leads to the highest computational complexity of the four discussed methods. Concatenating the question and the passage further increases the computational effort, which scales quadratically with the input sequence length due to the self-attention mechanism. Because of the missing cached representations, the cross-encoder is too slow in a real-world information retrieval scenario. In practice, the retrieval system can use it to rerank precomputed passages to enhance precision.

\section{Evaluation}
\label{sec:evaluation}
We evaluate our CAPR objective for the answer passage retrieval task in a zero-shot scenario on four healthcare datasets: WikiSectionQA (WS), MedQuAD (MQ), HealthQA (HQ) and MIMIC-III (MM-III) (see \autoref{tab:testdatastats}). For each dataset we use the plain section texts as passage candidates. We use Wikipedia or MIMIC-III as training data (see \autoref{tab:trainingdata_stats}). To highlight the properties of CAPR in different settings, we compare four architectures: CDV, Bi-encoder, Poly-encoder and Cross-encoder. For all models, we use either pretrained FastText word embeddings or a pretrained BioBERT model. 

\begin{table}[t]
\centering
 \begin{tabular}{l@{\hspace{0.1cm}} r@{\hspace{0.3cm}} r@{\hspace{0.3cm}} r@{\hspace{0.3cm}} r} 
 \hline
    & WS & MQ & HQ & MM-III \\
    \hline
    \# docs         & 716    & 171    & 1,111   & 7,465     \\
    \# passages     & 4,367  & 1,060  & 3,762   & 213,788   \\
    avg sents/doc   & 64     & 83     & 47      & 538     \\
    avg sents/psg.  & 10     & 13     & 14      & 19      \\
    avg tkns/psg.   & 298    & 292    & 291     & 163     \\
    avg tkns/sent.  & 30     & 24     & 23      & 11      \\
    labels          & meta   & meta   & human   & extracted  \\
\hline
\end{tabular}
\caption{Properties of the evaluation datasets. Entity and aspect labels are either provided by human annotators, metadata or were automatically extracted.}
\label{tab:testdatastats}
\end{table}

\subsection{Datasets for Healthcare Answer Retrieval}
\begin{table}[t]
\centering
 \begin{tabular}{l r r} 
    \hline
    & Wikipedia & MIMIC-III\\
    \hline
    \# docs & 8,597 & 25,202\\
    \# passages & 51,299 & 55,189 \\
    avg sents/doc & 53 & 80 \\
    avg sents/psg. & 7 & 36\\
    avg tkns/psg. & 186 & 448\\
    avg tkns/sent. & 31 & 13\\
 \hline
 \end{tabular}
 \caption{Properties of the training datasets. Entity and aspect labels are generated using CAPR.}
 \label{tab:trainingdata_stats}
\end{table}

\begin{table*}[h!]
    \centering
    \begin{tabular}{p{4.5cm}  r r  r r  r r  r r }
    \hline
    \textbf{Model} & \multicolumn{2}{c}{\textbf{WikiSection}} & \multicolumn{2}{c}{\textbf{MedQuAD}} & \multicolumn{2}{c}{\textbf{HealthQA}} & \multicolumn{2}{c}{\textbf{MIMIC-III}} \\
     & R@1 & R@5 & R@1 & R@5 & R@1 & R@5 & R@1 & R@5 \\
    TF-IDF                                  & 17.60 & 52.63 & 24.87 & 72.18 & 18.22 & 55.06 & 0.64  & 0.65 \\
    BM25                                    & 24.09 & 59.89 & 31.98 & 78.20 & 23.37 & 56.46 & 4.93  & 4.97 \\
    \hline
    \multicolumn{9}{l}{\textbf{Trained on MIMIC-III}} \\
    HAR                                     & 1.37  & 7.15  & 1.71  & 6.10  & 1.26  & 5.49  & 48.15 & 69.29 \\
    CDV+FastText                            & 1.90  & 10.68 & 3.00  & 12.57 & 2.77  & 9.78  & 47.80 & 59.57 \\
    CDV+BioBERT                             & 1.15  & 7.78  & 1.70  & 9.97  & 1.30  & 7.30  & 52.83 & 77.63 \\
    CDV+BioBERT-finetune                    & 2.43  & 8.57  & 4.31  & 12.00 & 2.46  & 10.06 & 56.77 & 81.05 \\
    CAPR-BioBERT-Bi-shared                  & 17.40 & 49.80 & 27.03 & 56.19 & 23.47 & 60.10 & 46.60 & 52.40 \\
    CAPR-BioBERT-Bi                         & 20.14 & 54.65 & 34.00 & 64.8  & 25.00 & 59.02 & 47.29 & 56.82 \\
    CAPR-BioBERT-Poly                       & 4.57  & 19.32 & 5.95  & 23.85 & 7.32  & 23.50 & 65.56 & 85.39 \\
    CAPR-BioBERT-Cross                      & \textbf{36.88} & \textbf{69.71} & \textbf{56.22} & \textbf{86.50} & \textbf{36.88} & \textbf{72.07} & \textbf{71.96} & \textbf{ 85.61} \\
    CAPR-Longformer-Bi                      & 2.35  & 12.10 & 4.56  & 16.45 & 4.36  & 16.82 & 3.19  & 10.29\\
    CAPR-Longformer-Poly                    & 2.54  & 11.30 & 3.70  & 13.30 & 3.17  & 10.09 & 54.97 & 76.00 \\
    \hline
    \multicolumn{9}{l}{\textbf{Trained on Wikipedia}} \\
    HAR                                     & 45.31 & 74.17 & 55.65 & 85.45 & 43.20 & 76.53 & 26.29 & 69.40\\      
    CDV+FastText                            & 60.61 & 93.87 & 39.60 & 79.00 & 36.04 & 64.92 & 5.39 & 16.59 \\
    CDV+BioBERT                             & \textbf{64.50} & 93.6 & 44.31 & 72.55 & 40.96 & 61.12 & 4.60 & 12.20 \\
    CDV+BioBERT-finetune                    & 63.39 & \textbf{95.47} & 43.09 & 79.18 & 40.59 & 76.62 & 1.61 & 8.22 \\
    CAPR-BioBERT-Bi-shared                  & 51.00 & 85.10 & 64.40 & 91.79 & 53.56 & 86.05 & 48.80 & 58.12 \\
    CAPR-BioBERT-Bi                         & 54.72 & 87.19 & 62.98 & 89.80 & 56.60 & 87.45 & 48.76 & 57.10 \\
    CAPR-BERT-base-Bi                       & 35.08 & 68.70 & 46.74 & 76.82 & 41.68 & 75.33 & 48.95 & 62.52 \\
    CAPR-BioBERT-Poly                       & 48.70 & 58.60 & 63.30 & 94.80 & 52.70 & 86.60 & 52.30 & 64.40 \\
    CAPR-BioBERT-Cross                      & 60.82 & 90.23 & \textbf{77.90} & \textbf{97.95} & \textbf{62.82} & \textbf{91.12} & \textbf{64.89} & \textbf{78.81} \\
    CAPR-Longformer-Bi                      & 49.40  & 86.88  & 49.77  & 88.84 & 41.78 & 80.57 & 21.02 & 45.96 \\
    CAPR-Longformer-Poly                    & 16.50 & 67.07 & 25.30 & 78.63 & 17.92 & 65.53 & 18.24 & 46.06\\
    \hline
    \multicolumn{9}{l}{\textbf{Trained on Wikipedia and MIMIC-III}} \\
    CDV+BioBERT                             & 57.09 & 93.73 & 37.11 & 77.96 & 35.64 & 65.60 & 51.59 & 79.45 \\
    CDV+BioBERT-finetune                    & \textbf{61.71} & \textbf{95.07} & 40.45 & 79.65 & 40.10 & 68.14 & 57.23 & 81.27 \\
    CAPR-BioBERT-Bi                         & 50.45 & 84.53 & \textbf{64.44} & \textbf{93.59} & 51.58 & 84.94 & 52.35 & 66.23 \\
    CAPR-BioBERT-Poly                       & 47.68 & 84.10 & 61.10 & 92.15 &\textbf{51.88}& \textbf{85.11} & 71.60 & 86.89 \\
    CAPR-BioBERT-Cross                      & 42.29 & 81.47 & 58.55 & 93.43 & 43.96 & 81.80 & \textbf{72.93} & \textbf{86.79} \\
    \hline
    \end{tabular}
    \caption{Results for cross-domain passage retrieval. Scores are given in \% Recall@\{1,5\}.}
    \label{tab:quantitative_results}
\end{table*}

\paragraph{MIMIC-III clinical notes.}
Our novel passage retrieval dataset is based on clinical admission discharge notes. MIMIC-III (Medical Information Mart for Intensive Care III) \cite{mimiciii} is a database of de-identified health-related data, associated with over 40,000 patients who stayed in critical care units of the Beth Israel Deaconess Medical Center between 2001 and 2002. We split \textbf{25,202} documents as training and \textbf{7,465} documents as evaluation dataset. In contrast to the Wikipedia dataset, MIMIC-III contains much more and longer documents with higher variance.

\paragraph{WikiSectionQA}
\cite{arnold2020learning} consists of 716 disease-related articles from Wikipedia. It uses the hyperlink-references and Wikidata IDs from entities, section headings from articles and provides both entity and aspect annotations for each sentence in each article.

\paragraph{MedQuAD}
\cite{abacha2019question} consists of 47,457 medical question-answer pairs from multiple trusted sources of the National Institutes of Health (NIH), such as National Cancer Institute, Genetic and Rare Diseases, National Institute of Diabetes and Digestive and Kidney Diseases and the National Institute of Neurological Disorders and Stroke. We use the test split of 1,060 queries introduced by \citet{arnold2020learning}.

\paragraph{HealthQA}
\cite{zhu2019hierarchical} consists of 7,517 questions and 7,355 documents (article-sections) collected from the health services website Patient\footnote{\href{https://patient.info/}{https://patient.info/}} which were manually labeled by clinical experts.

\subsection{Evaluation Setup}

\paragraph{Candidate passages.} In order to enable a fair comparison of the effectiveness of all retrieval methods on large datasets, we use a re-ranking approach. We exactly follow the setup of \citet{arnold2020learning} and select 64 potentially relevant passages for each query using BM25 (WS, MQ and HQ) or random sampling (MIMIC-III), which are then scored using the proposed methods.

\paragraph{Ranking Methods.} 
For all experiments, we provide a baseline score using term-based methods TF-IDF \cite{sparck1972statistical} and BM25 \cite{robertson1995okapi}, which are commonly used in full-text search engines. As another strong baseline, we consider Hierarchical Attention Retrieval (HAR) \cite{zhu2019hierarchical}, which outperforms all models in the MatchZoo benchmark \cite{guo2019matchzoo} on multiple medical domain retrieval tasks, as recently shown by \citet{zhu2019hierarchical} and \citet{arnold2020learning}. In addition, we report state-of-the-art scores for CDV \cite{arnold2020learning} and reproduced their scores using a model fine-tuned to our training data and averaging the cosine similarity of each sentence embedding in a passage.
We score the passages in CAPR (Bi- and Poly-encoder) with a nearest neighbor search using cosine similarity between the query and all passage candidates. In the case of the Cross-encoder, we rank the passage with respect to the query by the predicted score. As base model for CAPR, we compare BERT \cite{devlin2019bert}, BioBERT \cite{lee2019biobert} and Longformer \cite{Beltagy2020Longformer}.

\paragraph{Evaluation Metrics}
We report the Recall of the top ranked passage (R@1) and top-5 Recall (R@5), which best represents a clinical information retrieval scenario where a doctor wants to obtain information from multiple possible responses.

\subsection{Implementation Details}
We implement our models with PyTorch and the HuggingFace
transformer library \cite{wolf2019huggingface}. We manually chose the hyperparameters for all models. We use the AdamW optimizer, a learning-rate of $2\cdot10^-5$ and a batchsize of 32. We trained all models for 50 epochs in 16 bit half precision on a single Nvidia V100.

\subsection{Experimental Results}
The evaluation results are shown in \autoref{tab:quantitative_results}.
\paragraph{Transformer-based architectures drastically outperform CDV and HAR.}
Transformer based architectures with Bi-, Poly- and Cross-encoder generalize much better across the diseases domain. In contrast, CDV and HAR do not generalize to our clinical domain and only provide decent results when being trained on clinical data.
The generalization of HAR and CDV improves when being trained on both datasets together, especially in the clinical domain.

\paragraph{Domain-adapted outperform General Language Models.}
\citet{gururangan2020dontstop} observe that pretrained language models in special domains outperform common domain language models. We observe similar behaviour with BERT base and the Longformer that perform worse than BioBERT on disease related datasets. The Longformer outperforms BERT base on the Wikipedia datasets and performs even better than BioBERT in the Poly-encoder trained on MIMIC-III. This suggests that while in-domain pretraining of the language model is a key factor for good retrieval performance, the length of the processable sequence may play also an important role.

\paragraph{Query-Passage interaction is important.}
The Poly- and Cross-encoders typically perform best on all datasets. This implies that interaction between query and passage features is important for the retrieval. The Cross-encoder performs best when being trained in isolation on Wikipedia or MIMIC-III. If we train the Cross-encoder on both datasets together, we observe a significant performance loss on WikiQA, MedQuAD and HealthQA. 

\paragraph{CDV Fine-Tuning is not effective.}
Fine-Tuning the sentence encoder of the CDV architecture only slightly improves the performance of the model. This suggests that the sentence embedding via average or CLS pooling already contains most of the features required to predict the entity or aspect embedding. We find that the fine-tuned model performs roughly the same as the baseline, improving scores by only 4\% on MIMIC-III.

\paragraph{Retrieval latency: Poly-encoder is the optimal choice.}
\begin{figure}[t]
    \includegraphics[width=\linewidth]{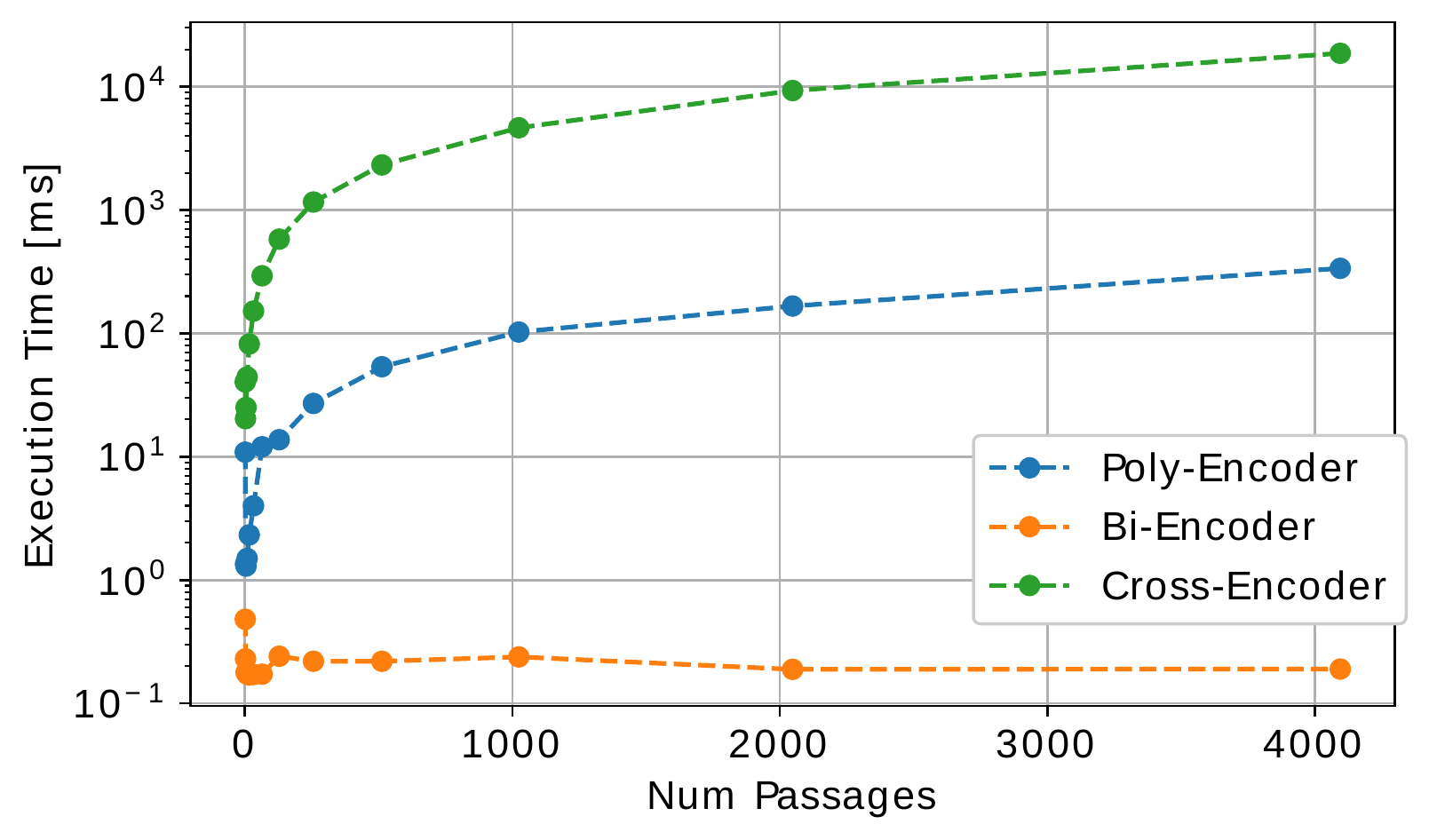}
    \caption{Retrieval latency of the Bi-, Poly- and Cross-encoder for a single query on an increasing number of passages.}
    \label{fig:latency_plot}
\end{figure}
We evaluate the retrieval latency for the Bi-, Poly- and Cross-encoder with a passage length of 512 tokens on a single Nvidia V100 GPU.
The Cross-encoder requires around ten seconds for a single query on 2,048 passages. The Bi- and Poly-encoders manage to retrieve the passage in under a second.
However, performance optimization techniques, such as quantization \cite{zafirq8bert2019}, knowledge distillation \cite{Xu-2020-Theseus} or weight pruning \cite{gordon-2020-compressingbert}, can further increase the performance, especially regarding the Cross-encoder. As expected, the Poly-encoder is slower than the Bi-encoder due to the additional attention layer. However, it provides more accurate results.

\subsection{Qualitative Error Analysis}
In our error analysis for each model and data set, we observe 50 randomly chosen mismatched samples to spot erroneous behaviour for each model. 

\paragraph{Anonymized sections.}
In our pretraining, we assume that the query and the passage contain enough information to learn the relationship between them. 
MIMIC-III contains anonymized names, dates or locations. Due to our automatic labeling process, our dataset contains many passages-query pairs with mostly anonymized information in the passage.
We observe that pretraining models on Wikipedia leads to in average 45\% of erroneous samples. This number reduces to 8\% if the models are pretrained on MIMIC-III.

\paragraph{Long documents and passages.}
WikiSectionQA, MedQuAD and HealthQA consist mostly of short passages (see \autoref{tab:testdatastats}). We observe on average 14.4\% of mismatched passages that are longer than 512 tokens, exceeding BERT's maximum sequence length. By its nature, MIMIC-III contains much longer passages where we observe 42.6\% of mismatched queries that exceed the maximum sequence length. 
We truncate very long passages to 512 tokens. This can lead to errors if the requested entity is discussed after the token limit. We observe better performance for the Longformer on MIMIC-III, indicating that it benefits from long range context dependencies, although it was not fine-tuned on medical domain data. We consider recent approaches such as BigBird \cite{zaheer2020bigbird} or Reformer \cite{kitaev2020reformer}, which can encode longer sequences, as future work.


\paragraph{Mismatched entity and aspect subclasses.}
We observe that most mismatches contain at least a related entity (57.8\%) or a related aspect (21.3\%) in the case of Wikipedia, MedQuAD and HealthQA. In the case of the MIMIC-III evaluation dataset, the ratio of related aspects of mismatches is 59.3\% and related entities 32.7\%. Related aspects occur due to hierarchical information, such as subclasses of diseases or negations. We observe that NER preprocessing excludes certain negotiations and thus creates wrong labels. Therefore, we find queries with the entity \textit{nausea} that link to passages containing: \textit{double vision, no nausea or vomiting}. 

\paragraph{Structural vs. topical aspects.}~
Many labels only contain structural information content in the MIMIC-III dataset. Due to the automatic labeling, we rely on section headings as aspect annotations for the dataset. 
Those section headings typically are structural, such as \textit{history of present illness} or \textit{laboratory results}. Therefore, they do not contain any specific patient-related information and only serve the purpose of structuring the document.
Since we automatically annotate the document based on these headings, the MIMIC-III based dataset contains many queries that have little information content. In contrast, Wikipedia headings on disease-related articles are more topical (e.g. \textit{Symptoms} or \textit{Diagnostics}).

\paragraph{Note-like structure in MIMIC-III.}
MIMIC-III differs drastically in its structure in comparison to Wikipedia, MedQuAD and HealthQA. The clinical notes contain consistent information in single sentences and often do not span multiple sentences or paragraphs. This makes the task of finding a single passage for a relevant entity and aspect much more difficult, because one passage usually contains several important entities.


\section{Conclusion and Future Work}
\label{sec:conclusion}
Many domain-specific scenarios, such as retrieving clinical notes, only have limited access to labeled training data. We solve this problem with CAPR using self-supervised training and investigate four powerful query-passage-encoder architectures. From our extensive evaluation on four datasets annotated by metadata or human experts, we report that CAPR outperforms existing supervised methods and can answer ad-hoc queries with short latency. 

Our future work envisions more complex question types and methods to improve entity and topical aspect recognition. For the latter, we propose to investigate the combination of multiple NER models for higher recall of recognized entities during training.
Furthermore, future research should be devoted to the impacts of the application of recent Transformer architectures that are able to capture longer sequences such as the Reformer \cite{kitaev2020reformer} or Big Bird \cite{zaheer2020bigbird}.


\bibliographystyle{acl_natbib}
\bibliography{anthology,acl2021}


\end{document}